\documentstyle[emulateapj,psfig,natbib,amsfonts,amsmath,graphicx]{article}
\def\swift{{\it Swift}}
\def\nod{\nodata}

\def\ociw{1}
\def\prince{2}
\def\hubble{3}

\begin{document}

\title{The Prompt Gamma-Ray and Afterglow Energies of Short-Duration
Gamma-Ray Bursts}

\author{
E.~Berger\altaffilmark{\ociw,}\altaffilmark{\prince,}\altaffilmark{\hubble}
}

\altaffiltext{\ociw}{Observatories of the Carnegie Institution
of Washington, 813 Santa Barbara Street, Pasadena, CA 91101}

\altaffiltext{\prince}{Princeton University Observatory, Peyton 
Hall, Ivy Lane, Princeton, NJ 08544}

\altaffiltext{\hubble}{Hubble Fellow}

\begin{abstract} 
I present an analysis of the $\gamma$-ray and afterglow energies of
the complete sample of 17 short-duration GRBs with prompt X-ray
follow-up.  I find that $80\%$ of the bursts exhibit a linear
correlation between their $\gamma$-ray fluence and the afterglow X-ray
flux normalized to $t=1$ d, a proxy for the kinetic energy of the
blast wave ($F_{X,1}\propto F_\gamma^{1.01\pm 0.09}$).  An even
tighter correlation is evident between $E_{\gamma,{\rm iso}}$ and
$L_{X,1}$ for the subset of 13 bursts with measured or constrained
redshifts.  The remaining $20\%$ of the bursts have values of
$F_{X,1}/F_\gamma$ that are suppressed by about three orders of
magnitude, likely because of low circumburst densities \citep{nak07}.
These results have several important implications: (i) The X-ray
luminosity is generally a robust proxy for the blast wave kinetic
energy, indicating $\nu_X>\nu_c$ and hence a circumburst density
$n\gtrsim 0.05$ cm$^{-3}$; (ii) most short GRBs have a narrow range of
$\gamma$-ray efficiency, with $\langle\epsilon_\gamma\rangle\approx
0.85$ and a spread of 0.14 dex; and (iii) the isotropic-equivalent
energies span $10^{48}-10^{52}$ erg.  Furthermore, I find tentative
evidence for jet collimation in the two bursts with the highest
$E_{\gamma,{\rm iso}}$, perhaps indicative of the same inverse
correlation that leads to a narrow distribution of true energies in
long GRBs.  I find no clear evidence for a relation between the
overall energy release and host galaxy type, but a positive
correlation with duration may be present, albeit with a large scatter.
Finally, I note that the outlier fraction of $20\%$ is similar to the
proposed fraction of short GRBs from dynamically-formed neutron star
binaries in globular clusters.  This scenario may naturally explain
the bimodality of the $F_X/F_\gamma$ distribution and the low
circumburst densities without invoking speculative kick velocities of
several hundred km s$^{-1}$.
\end{abstract}
 

\section{Introduction}
\label{sec:intro}

One of the fundamental parameters of any explosive phenomenon is the
overall energy release.  In the specific case of gamma-ray bursts
(GRBs), it is essential to measure the energy of both the prompt
$\gamma$-ray phase ($E_\gamma$) and the blast wave which powers the
afterglow ($E_K$), since these two quantities define the total
relativistic output of the central engine ($E_{\rm rel}$) and the
efficiency of energy dissipation in the form of $\gamma$-rays
($\epsilon_\gamma\equiv E_\gamma/E_{\rm rel}$).  While the
isotropic-equivalent prompt energy is easily derived from the observed
fluence ($F_\gamma$) and distance ($d_L(z)$) of the burst, a
measurement of $E_K$ is more challenging because it requires
observations of the afterglow over a broad range in time and frequency
(e.g., \citealt{bsf+00,pk02,yhs+03}).  Luckily, the afterglow flux at
frequencies above the synchrotron cooling frequency, $\nu_c$, provides
a robust proxy for $E_K$ since it is independent of the circumburst
density and depends only weakly on the fraction of shock energy in
magnetic fields, $\epsilon_B$
\citep{kum00,fw01,bkf03}.  For typical parameters \citep{pk02,bkf03},
the soft X-ray band lies above $\nu_c$, and therefore the X-ray
luminosity, $L_X\propto\epsilon_eE_K$; here $\epsilon_e$ is the
fraction of shock energy in relativistic electrons.

In the case of the long-duration GRBs ($T_{90}\gtrsim 2$ s), extensive
studies of the relativistic energy output based on these various
techniques suggest that for most bursts $E_\gamma$, $E_K$, and most
importantly their sum, span less than an order of magnitude around
$\sim 10^{51}$ erg {\it when corrected for jet collimation of the
ejecta} \citep{fks+01,pk02,bkf03,bfk03,bkp+03}.  Recent observations
of low redshift long GRBs paint a more complicated picture in which a
population of sub-energetic bursts ($\sim 10^{50}$ erg) likely
dominates the overall GRB rate \citep{skb+04,skn+06}.  Still, the
conclusion that the energy release of long GRBs is in the range
typical of supernovae, and that they are collimated with opening
angles of $\theta_j\sim 5^\circ$, has placed valuable constraints on
the progenitor and engine models.

The discovery of afterglow emission from short-duration GRBs less than
2 years ago makes it now possible to carry out a similar analysis of 
their total energy output.  Early observations indicated a lower
isotropic-equivalent energy release compared to long GRBs, $\sim
10^{48}-10^{49}$ erg, in both the prompt and afterglow phase
\citep{bpc+05,ffp+05,gso+05,bpp+06}, as well as wider jet opening
angles \citep{bgc+06,sbk+06}.  In addition, \citet{nak07} noted based
on a small sample of six bursts that the ratio of afterglow X-ray flux
and $\gamma$-ray fluence, $F_{X}t/F_\gamma$, appears to have a bimodal
distribution with values of $\sim 10^{-2}$ and $\lesssim 10^{-4}$.
They argue that the latter bursts require a low circumburst density,
$n\lesssim 10^{-5}$ cm$^{-3}$, likely due to ejection of the
progenitor system from its host galaxy into the intergalactic
medium\footnotemark\footnotetext{\citet{nak07} also propose that a
wide dispersion in the shock parameters or $\epsilon_\gamma$ may lead
to the bimodal distribution.}.

The recent association of several short GRBs with faint galaxies, of
which several have been spectroscopically confirmed to be at $z\sim
0.4-1.1$, indicates a wide redshift distribution, and hence a wider
range of energies than previously proposed \citep{bfp+06}.  Here, I
take advantage of these new observations to study the prompt and
afterglow energies of the complete sample of 17 short bursts with
prompt X-ray follow-up, a much larger sample than previously
investigated.  Of these 17 bursts, eight have measured redshifts, and
an additional five have approximate or constrained redshifts.  Using
the $\gamma$-ray fluences and X-ray fluxes normalized to 1 day after
the burst I show that $E_{\gamma,{\rm iso}}$ and $E_{K,{\rm iso}}$
span at least three orders of magnitude (possibly up to $\sim 10^{52}$
erg).  More importantly, I establish that $80\%$ of the bursts exhibit
a tight correlation between their prompt and afterglow energy output,
suggesting a narrow distribution of $\epsilon_\gamma$, while $20\%$
indeed have ratios of $F_{X}/F_\gamma$ that are suppressed by three
orders of magnitude.  These results place important constraints on the
progenitor and energy extraction models, as well as the properties of
the circumburst medium and the shock microphysical parameters.

\section{Prompt Gamma-Ray and X-ray Afterglow Energies}
\label{sec:data}

The $\gamma$-ray and X-ray afterglow properties of the 17 short bursts
with prompt X-ray follow-up are listed in Table~\ref{tab:data}.  I
also list the available measurements or constraints on the redshift
from host galaxy and afterglow observations.  The relevant references
are provided in the Table.

For GRB\,061210 the X-ray fluxes have not been published and I
therefore obtained the XRT data from the High Energy Astrophysics
Science Archive Research
Center\footnote{http://heasarc.gsfc.nasa.gov/W3Browse/}.  I processed
the data with the {\tt xrtpipeline} script packaged within the HEAsoft
software, using the default grade selection and screening parameters
to produce a light curve for the $0.3-10$ keV energy range.  Using a
Galactic neutral hydrogen column density, $N_H\approx 3.4\times
10^{20}$ cm$^{-2}$, and a typical $\beta_X\approx -1.1$, I find a
conversion rate of 1 count s$^{-1}=4\times 10^{-11}$ erg cm$^{-2}$
s$^{-1}$.  In addition, inspection of the X-ray light curve of
GRB\,061006 \citep{sbp+06} reveals a possible steepening at $t\sim
10^5$ s.  To investigate this possibility I obtained the XRT data from
HEASARC and followed the same analysis procedure as for GRB\,061210.
I find a likely steepening using data that were obtained $3-4$ days
after the burst, with a flux decay rate, $\alpha_X\approx -1.9$.

In the analysis below I use the standard cosmological parameters:
$H_0=70$ km s$^{-1}$ Mpc$^{-1}$, $\Omega_m=0.27$, and
$\Omega_\Lambda=0.73$.

\begin{figure}
\epsscale{0.8}
\plotone{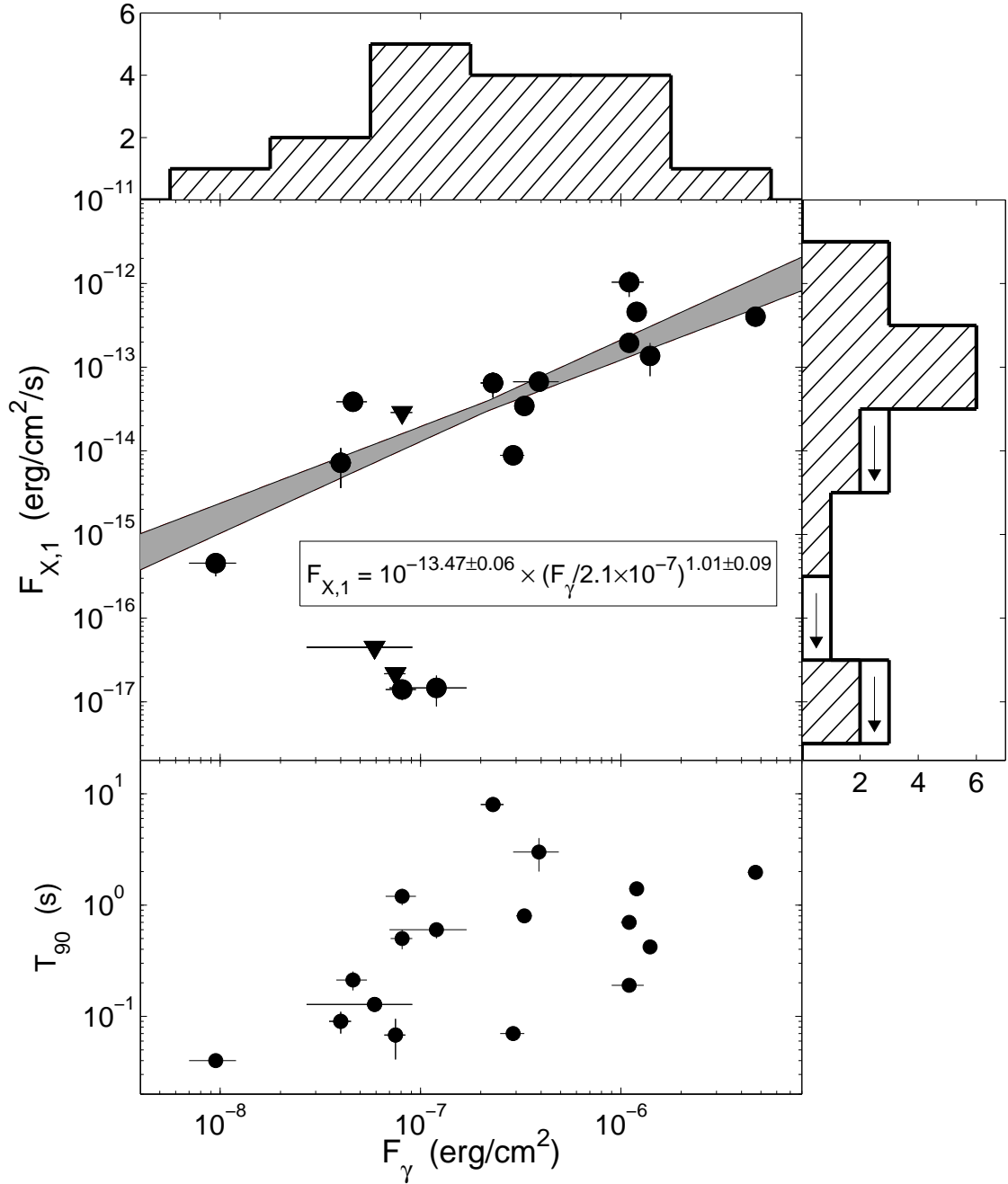}
\caption{X-ray flux normalized to $t=1$ d and burst duration
($T_{90}$) plotted against the $\gamma$-ray fluence for the complete
sample of 17 short GRBs with prompt XRT follow-up.  The majority of
the bursts ($80\%$) follow a clear linear relation between $F_{X,1}$
and $F_\gamma$ suggesting a relatively standard fraction of energy
emitted in $\gamma$-rays.  At the same time, a subset of the short
bursts have X-ray fluxes that are suppressed by about three orders of
magnitude.  The projected histograms indicate the full range of values
for each observed quantity (arrows designate upper limits on the X-ray
flux).  There is also a possible correlation between $T_{90}$ and
$F_\gamma$ (with a large dispersion) -- the null hypothesis of no
correlation has a probability of only 0.75\%.
\label{fig:fxfgamma}}
\end{figure}

In Figure~\ref{fig:fxfgamma}, I plot the histograms of $F_\gamma$ and
$F_{X,1}$, the X-ray flux normalized to $t=1$ d after the burst.  I
use the observed values of $\alpha_X$ when available, or the median
value, $\langle\alpha_X\rangle\approx -1.1$, when the decay rate is
not known.  The majority of the bursts span about two orders of
magnitude around $F_{X,1}\sim 10^{-13}$ erg cm$^{-2}$ $s^{-1}$,
similar to the width of the distribution for long GRBs
\citep{bkf03,bkf+05}.  However, there is a noticeable tail extending
to fainter fluxes, which leads to an overall spread of about five
orders of magnitude and a median flux of $3\times 10^{-14}$ erg
cm$^{-2}$ s$^{-1}$.  The fluence distribution is narrower, with an
overall width of about 2.7 orders of magnitude around a median value
of $2\times 10^{-7}$ erg cm$^{-1}$.  The median values of both
$F_\gamma$ and $F_{X,1}$ are about a factor of five smaller than for
\swift\ long GRBs \citep{bkf+05}.

More interesting is the apparent correlation between $F_\gamma$ and
$F_{X,1}$ for the majority of the bursts (Figure~\ref{fig:fxfgamma}).
The Spearman's rank correlation coefficient is $\rho=0.86$, indicating
a null hypothesis (no correlation) probability of only $5.3\times
10^{-5}$.  Fitting the data, I find that the best-fit relation is
linear, $F_{X,1}= 10^{-13.47\pm 0.06}\times (F_\gamma/
10^{-6.68})^{1.01\pm 0.09}$ erg cm$^{-2}$ s$^{-1}$ with a scatter of
about $0.4$ dex based on a fit to 13 of the 17 bursts that follow the
trend.  The tight linear correlation suggests that $\epsilon_\gamma$
has a narrow distribution, and that in most bursts the X-ray band
indeed lies above the cooling frequency and traces $E_K$.  Based on
the observed correlation I infer that $\epsilon_eE_K/ E_\gamma\propto
t_1F_{X,1}/F_\gamma\approx 0.016$ (see also \citealt{nak07}).  For a
typical $\epsilon_e\sim 0.1$, the fraction of energy emitted in
$\gamma$-rays is thus $\epsilon_\gamma\sim 85\%$, with an overall
range of about $55-95\%$, similar to the values inferred for long GRBs
\citep{pk02}.

\begin{figure}
\epsscale{0.8}
\plotone{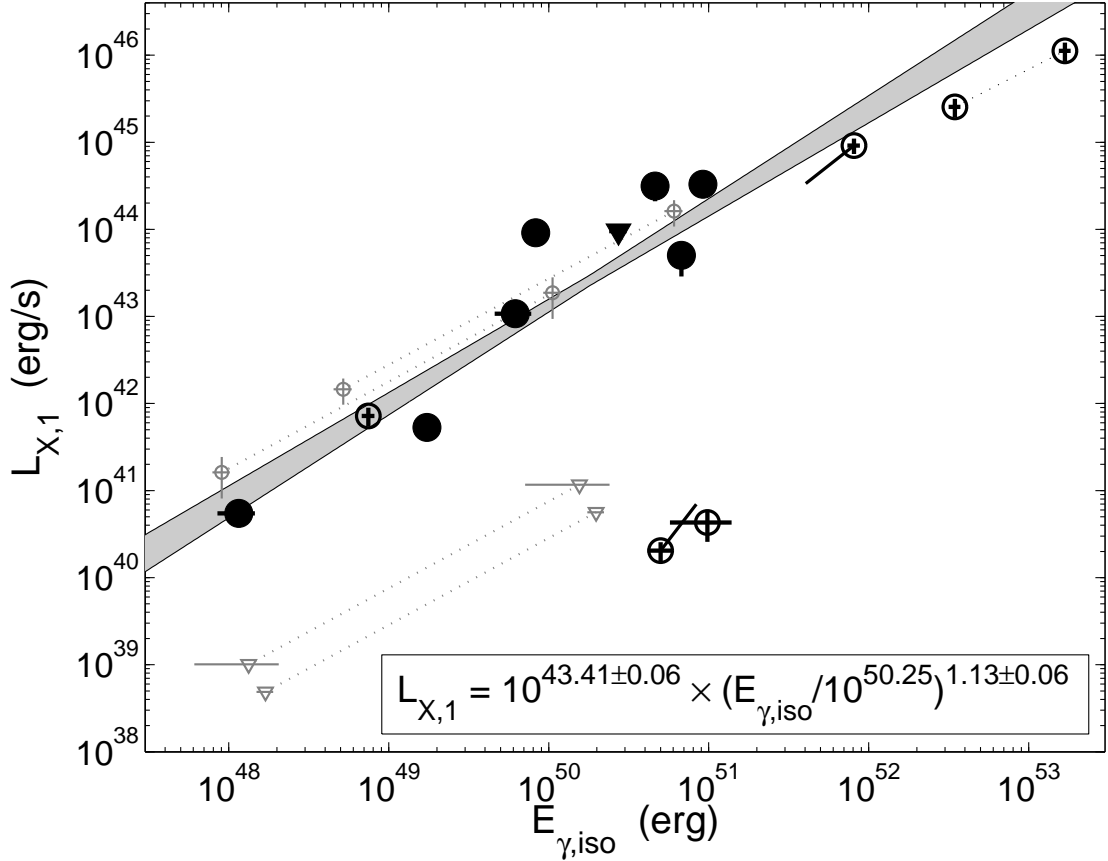}
\caption{X-ray luminosity normalized to $t=1$ d (a proxy for
$\epsilon_eE_{K,{\rm iso}}$) plotted against $E_{\gamma,{\rm iso}}$
for the short bursts with a known redshift (solid black circles),
redshift constraints (open black circles) and without any redshift
information (gray symbols connected by dotted lines).  For the latter
I plot the inferred values at $z=0.1$ and $z=1$, corresponding roughly
to the lowest and highest redshifts securely measured to date.  The
isotropic-equivalent relativistic energies are as least as high as
$10^{51}$ erg, and may approach ${\rm few}\times 10^{52}$ erg for some
short bursts (see also \citealt{bfp+06}).
\label{fig:egammalx}}
\end{figure}

An even tighter correlation is evident between the X-ray luminosity,
$L_{X,1}\propto\epsilon_eE_{K,{\rm iso}}$, and $E_{\gamma,{\rm iso}}$;
Figure~\ref{fig:egammalx}.  The isotropic X-ray luminosity is given
by:
\begin{equation}
L_{X,1}=4\pi d_L^2 F_{X,1}(1+z)^{\alpha_X-\beta_X-1},
\label{eqn:lx}
\end{equation}
where for $\nu_X>\nu_c$, $\alpha_X-\beta_X=(2-p)/4$, $p\sim 2-3$ is
the power law index of the electron distribution, $N(\gamma)\propto
\gamma^{-p}$, and $d_L$ is the luminosity distance.  The isotropic
$\gamma$-ray energy (neglecting redshift-dependent bolometric 
corrections) is given by:
\begin{equation}
E_{\gamma,{\rm iso}}=4\pi d_L^2 F_\gamma (1+z)^{-1}.
\label{eqn:egamma}
\end{equation}

To assess $L_{X,1}$ and $E_{\gamma,{\rm iso}}$ I use the redshifts
measured for eight bursts from their host galaxies, as well as
approximate redshifts for three bursts (050813, 060121, and 061201)
and limits for an additional two (051210 and 060313).  The Spearman's
rank correlation coefficient is $\rho=0.94$, (null hypothesis
probability of only $3.2\times 10^{-6}$).  The correlation appears to
be slightly non-linear, $L_{X,1}\propto E_{\gamma,{\rm iso}}^{1.13\pm
0.06}$, primarily because for $p>2$ the dependence on $(1+z)$ does not
cancel out when taking the ratio of Equations~\ref{eqn:lx} and
\ref{eqn:egamma}.  As can be seen from Table~\ref{tab:data}, the
bolometric correction factors to $E_{\gamma,{\rm iso}}$ are in the
range of $\sim 3-15$ when converting from the $15-150$ keV band to the
$20-2000$ keV band.  These upward corrections to $E_{\gamma,{\rm
iso}}$ will tend to linearize the correlation.

In addition to the overall strong correlation, it is also clear from
Figure~\ref{fig:egammalx} that the range of inferred energies spans
$E_{\gamma,{\rm iso}}\sim 10^{48}-10^{51}$ erg ($15-150$ keV), and may
exceed $\sim 10^{52}$ erg if some of the bursts with unknown redshifts
are located at $z\gtrsim 1$, as suggested by their faint host galaxies
($R\gtrsim 23$ mag; \citealt{bfp+06}).  The full range of $3-4$ orders
of magnitude is similar to the spread in $E_{\gamma,{\rm iso}}$
observed for long GRBs \citep{fks+01,bfk03}.

\section{Implications} 
\label{sec:disc}

\subsection{A Narrow Distribution of $\epsilon_\gamma$}
\label{sec:epsilon}

The data presented in this paper conclusively show that the majority
($80\%$) of all short GRBs exhibit a tight correlation between their
isotropic-equivalent prompt $\gamma$-ray and blast wave kinetic
energies (see also \citealt{nak07}).  The inferred $\gamma$-ray
efficiency is $\sim 85\%$ with a narrow spread of about $0.14$ dex.
This result is indeed verified by the three bursts for which detailed
afterglow observations are available, with derived $\epsilon_\gamma$
values of 0.8 (GRB\,050709; \citealt{ffp+05}), 0.2 (GRB\,050724;
\citealt{bpc+05}), and 0.65 (GRB\,051221a; \citealt{sbk+06}).

The remaining $20\%$ of short bursts have X-ray fluxes that are
suppressed by about three orders of magnitude compared to their
$\gamma$-ray fluence.  I stress that if this is indeed due to a low
circumburst density (\S\ref{sec:outliers}), which leads to
$\nu_X<\nu_c$, then observations of the afterglow at higher X-ray
energies (above $h\nu_c$) should recover the same narrow distribution
of $\epsilon_\gamma$ seen for the bulk of the population.

The observed correlation and the inferred narrow distribution of
$\epsilon_\gamma$ have several crucial implications for short GRB
progenitor models, the energy extraction mechanism, and burst
properties such as the circumburst density and shock microphysics.
The overall narrow spread in $\epsilon_\gamma$, and the median value,
are similar to those observed in long GRBs \citep{pk02}, suggesting
that the properties of the relativistic outflow are similar for long
and short bursts and are generally independent of the identity of the
progenitor or the circumburst environment.

In addition, the tight correlation between $\gamma$-ray and afterglow
energies indicates that for most bursts the underlying assumption that
$\nu_X>\nu_c$ is indeed correct.  Therefore the required circumburst
densities are \citep{gs02} $n\gtrsim 0.05\,(1+z)^{-1/2}
\epsilon_{B,-1}^{-3/2} E_{50}^{1/2}$ cm$^{-3}$, typical of interstellar
environments.  In the context of binary compact object progenitors
(NS-NS, NS-BH), the inferred densities indicate that the majority of
the progenitors do not experience kick velocities that are large
enough for ejection into the intergalactic medium and/or that the
merger timescales are short enough that the distance traveled is
$\lesssim 10$ kpc.  Finally, the fraction of energy in relativistic
electrons, $\epsilon_e\approx(1-\epsilon_\gamma)E_K/E_\gamma$, must
also have a narrow distribution, since both $\epsilon_\gamma$ and
$E_K/E_\gamma$ have a narrow spread.

\subsection{Isotropic-Equivalent Energies and Beaming Corrections}

While there is a clear correlation between $E_{\gamma,{\rm iso}}$ and
$E_{K,{\rm iso}}$, the overall spread in isotropic-equivalent energies
appears to be wider than for the beaming-corrected energies of long
GRBs, {\it unless the outflows of short GRBs with the highest energies
are strongly collimated}.  To date, only GRB\,051221a exhibits
evidence for significant beaming, with $f_b\equiv [1-{\rm
cos}(\theta_j)]\approx 7.5\times 10^{-3}$ \citep{bgc+06,sbk+06}, while
GRB\,050709 appears to have a wide jet with $f_b\approx 0.06$
\citep{ffp+05}.  In general short GRB jets appear to be wider than
those of long GRBs with $\theta_j \gtrsim 10^\circ$ compared to
$\langle\theta_j\rangle\approx 5^\circ$ \citep{sbk+06}.

In the larger sample presented here two additional bursts with known
redshifts (061006 and 061210) exhibit steep decays, $\alpha_X\sim -2$,
at late time, reminiscent of a post jet break evolution (for which the
expected value is $\alpha_X=-p$ with $p\gtrsim 2$; \citealt{sph99}).
In addition to GRB\,051221a these bursts also have the largest secure
values of $E_{\gamma,{\rm iso}}$.  In the case of GRB\,061006 there is
a clear break at $t\approx 1\times 10^5$ s, while for GRB\,061210 the
light curve is already in the rapid decay phase at the time of the
first observations, $t\approx 2\times 10^5$ s.  Using the conversion
from jet break time to opening angle
\citep{sph99}:
\begin{equation}
\theta_j=0.21\,t_{j,1}^{3/8}(1+z)^{-3/8}E_{\gamma,{\rm
iso},51}^{-1/8} (\epsilon_\gamma/0.8)^{1/8}n_0^{1/8},  
\label{eqn:jet}
\end{equation}
I find $\theta_j\approx 0.12$ for GRB\,061006 and $\theta_j\lesssim
0.16$ for GRB\,061210 ($n_0=10^{-2}$ cm$^{-3}$).  Thus, in the context
of jet breaks, the beaming-corrected energies are $E_\gamma\approx
5\times 10^{48}$ erg and $\lesssim 6\times 10^{48}$ erg, respectively.
These values are within a factor of few of $E_\gamma\approx (1-2)
\times 10^{49}$ erg inferred for GRB\,051221a \citep{sbk+06}, and 
$E_\gamma\approx 2.1\times 10^{48}$ inferred for for GRB\,050709
\citep{ffp+05}.

If the observed steep decays are indeed due to jets, this leads to a
similar inverse correlation between $E_{\rm iso}$ and beaming
correction that is observed in long GRBs \citep{fks+01,bkf03,bfk03},
and thus to a narrow distribution of $E_{\rm rel}\sim 10^{48}-10^{49}$
erg.  Models of the relativistic outflows from black hole accretion
systems relevant for short GRBs indicate $E_{\rm rel,iso}\sim
10^{49}-10^{52}$ with opening angles of $\sim 5-15^\circ$
\citep{jam+06}, in good agreement with the observations presented
here.  It is thus possible that short GRBs also exhibit a nearly
standard energy release, with an overall scale that is two orders of
magnitude smaller than that of long GRBs.

The proposed beaming corrections also impact the nature of the energy
extraction mechanism.  In particular, extraction via $\nu\bar{\nu}$
annihilation has been argued to provide at most $E_{\rm rel}\sim {\rm
few}\times 10^{48}$ erg \citep{rr02}, which may be sufficient if the
beaming corrections are valid.  On the other hand, if the observed
steep decays are not due to jets, then the required large energies
likely point to MHD processes such as the Blandford-Znajek mechanism
\citep{bz77,rrd03,sbk+06}, which can produce luminosities in excess of
$10^{52}$ erg s$^{-2}$.  These possibilities have to be assessed with
a larger sample of bursts, with particular attention to
multi-wavelength observations that can differentiate between bona-fide
jet breaks and other scenarios (e.g., energy or density variations).

\subsection{Correlations with Duration and Host Galaxy Type?}

I further investigate whether there is a correlation between the
energy release and burst duration ($T_{90}$);
Figure~\ref{fig:fxfgamma}.  For the full sample I find a Spearman's
rank correlation coefficient, $\rho=0.58$, or a null hypothesis
probability of only $7.5\times 10^{-3}$.  Removing GRB\,050509b, which
has the shortest duration and smallest fluence, I find $\rho=0.49$, or
a probability of $2.6\times 10^{-2}$ that the null hypothesis is
satisfied.  Thus, I conclude that with the present sample the
correlation between fluence and duration is suggestive, though not
conclusive.

A more physically meaningful test is whether $E_{\gamma,{\rm iso}}$ is
correlated with the rest-frame duration, $T_{90}/(1+z)$.  In this
case, I find that for the eight bursts with secure redshifts the null
hypothesis has a probability of $4.5\times 10^{-2}$, so an overall
correlation does not appear to be statistically meaningful.
Curiously, the sample of six bursts with secure redshifts and
late-type host galaxies has $\rho=0.94$ (null hypothesis probability
of only $6\times 10^{-4}$), but this result is likely due to small
number statistics.  To conclude, there is some tentative evidence for
a positive correlation between energy release and duration, but this
will have to be re-assessed with a larger sample.

Similarly, I check for a relation between energy release and host
galaxy type.  This is the case for Type Ia supernovae (SNe Ia), which
tend to have lower luminosities in early-type galaxies than in
late-type galaxies, likely due to a dependence on the progenitor ages
\citep{ggb+05,slp+06}.  Short GRBs also occur in both types of
galaxies so a similar trend may shed light on the progenitor
properties.  In the sample presented here, three bursts have been
localized to early-type galaxies (050509b, 050724, and 050813), while
six have secure late-type hosts (050709, 051221a, 060801, 061006,
061210, and 061217).  As can be seen from Table~\ref{tab:data} both
groups appear to span the full range of $\gamma$-ray energies and
X-ray luminosities, suggesting no clear correlation with host galaxy
type for the present sample.

\subsection{The Nature of the Outliers}
\label{sec:outliers}

Finally, I return to the $20\%$ of outliers with an unusually low
ratio of $F_{X,1}/F_\gamma$.  The possible nature of these objects has
been discussed by \citet{nak07} who noted several possibilities for
the suppressed X-ray flux, in particular a low circumburst density
($n\lesssim 10^{-5}$ cm$^{-3}$), typical of the intergalactic medium.
This is argued to support the idea of large kick velocities, $\gtrsim
10^2$ km s$^{-1}$, for some progenitors.  Here I simply note that
while all four outliers have fluences at the low end of the
distribution, $F_\gamma\lesssim 10^{-7}$ erg cm$^{-2}$
(Figure~\ref{fig:fxfgamma}), the two with redshift constraints (050813
and 051210) appear to reside at $z\gtrsim 1.5$ and therefore their
energies are at the high end of the overall distribution.  No other
parameters clearly distinguish the outliers from the bulk of the
population that has a narrow distribution of $\epsilon_\gamma$,
suggesting that indeed an extrinsic parameter such as the density is
responsible for their low X-ray fluxes.

In the context of a low density interpretation for these outliers, and
given the clearly bimodal distribution, I propose the following
intriguing possibility.  The observed fraction of 20\% is similar to
predictions of the fraction of short GRBs that may arise from
dynamically-formed neutron star binaries in globular clusters
($10-30\%$; \citealt{gpm06}).  This scenario may naturally explain the
required low densities since the intra-cluster medium of globular
clusters has a typical limit\footnotemark\footnotetext{However, a
substantial intra-cluster density, $n\sim 0.07$ cm$^{-3}$, has been
claimed for the globular cluster 47 Tucanae \citep{fkl+01}.} of
$n\lesssim {\rm few}\times 10^{-5}$ cm$^{-3}$ \citep{lse+06}, likely
as a result of gas stripping from frequent passages through the
Galactic disk.  This scenario also naturally explains the bimodal
distribution of $F_X/F_\gamma$ (and hence densities), since the bursts
occur in either interstellar environments with $n\gtrsim 0.05$
cm$^{-3}$ (\S\ref{sec:epsilon}) or in the low density environments of
globular clusters.  Thus, this alternative scenario removes the need
for speculative large kick velocities for the outliers \citep{nak07},
which in any case predicts a more uniform distribution of densities
and hence $F_X/F_\gamma$ than currently observed.

\section{Summary}

Based on prompt $\gamma$-ray and X-ray afterglow observations of a
sample of 17 short GRBs with rapid XRT follow-up I find the following
results:
\begin{enumerate}
\item The majority of short GRBs (80\%) follow a linear correlation 
between $E_\gamma$ and $E_K$, which indicates a narrow distribution of
$\gamma$-ray efficiency (0.14 dex) with $\langle\epsilon_\gamma\rangle
\sim 0.85$, similar to the values for long GRBs.
\item The observed correlation also indicates that generally $\nu_X>
\nu_c$ and therefore $n\gtrsim 0.05$ cm$^{-3}$, indicative of 
interstellar explosion sites.
\item The isotropic-equivalent energies span $10^{48}-10^{52}$ erg, 
similar in width to the distribution for long GRBs, but lower by about
two orders of magnitude.
\item Possible beaming corrections for several short bursts with the 
highest values of $E_{\gamma,{\rm iso}}$ may point to an inverse
correlation such that the true energy release has a narrow
distribution of $\sim 10^{48}-10^{49}$.
\item There is a possible weak correlation between $E_{\gamma,{\rm 
iso}}$ and burst duration; no clear relation is evident between energy
release and host galaxy type.
\item A small fraction of outliers (20\%), with $F_X/F_\gamma$ lower by 
three orders of magnitude, are likely the result of low circumburst
densities.  The low density, strong bimodality of the distribution,
and similarity in predicted rates, are naturally explained in a
globular cluster origin as opposed to large kick velocities.
\end{enumerate}

With these properties of short GRBs well-established the next step is
to assess the importance of beaming corrections, and hence the true
distribution of total relativistic output from the central engine.

\acknowledgements 
I thank Alicia Soderberg and Ehud Nakar for valuable discussion, and
acknowledge support by NASA through Hubble Fellowship grant
HST-01171.01 awarded by the Space Telescope Science Institute, which
is operated by AURA, Inc.~for NASA under contract NAS 5-26555.


\clearpage
\begin{deluxetable}{llllclll}
\tablecolumns{8}
\tabcolsep0.1in\footnotesize
\tablewidth{0pc}
\tablecaption{$\gamma$-Ray and X-ray Properties of Short GRBs
\label{tab:data}}
\tablehead {
\colhead {GRB}            &
\colhead {$z$}            &
\colhead {$T_{90}$}       &
\colhead {$F_\gamma\,^a$} &
\colhead {$\Delta t$}     &
\colhead {$F_X$}          &
\colhead {$\alpha_X$}     &
\colhead {Refs.}          \\
\colhead {}                &
\colhead {}                &
\colhead {(s)}             &
\colhead {(erg cm$^{-2}$)} &
\colhead {(s)}             &
\colhead {(erg cm$^{-2}$ s$^{-1}$)} &
\colhead {}                &
\colhead {}          
}
\startdata
050509b & $0.226$        & $0.040\pm 0.004$ & $(9.5\pm 2.5)\times 10^{-9}$     & $200$             & $(3.6\pm 1.1)\times 10^{-13}$ & $-1.1$ & 1--2      \\
050709  & $0.1606$       & $0.070\pm 0.010$ & $(3.0\pm 0.4)\times 10^{-7}\,^b$ & $2.2\times 10^5$  & $(3.5\pm 0.5)\times 10^{-15}$ & $-1.0$ & 3--5      \\
050724  & $0.257$        & $3.000\pm 1.000$ & $(3.9\pm 1.0)\times 10^{-7}$     & $6625$            & $(8.8\pm 2.1)\times 10^{-13}$ & $-1.0$ & 6--8      \\
050813  & $\sim 1.8$     & $0.600\pm 0.100$ & $(1.2\pm 0.5)\times 10^{-7}$     & $2100$            & $(3.0\pm 1.2)\times 10^{-14}$ & $-2.0$ & 9--11     \\
050906  & \nod           & $0.128\pm 0.016$ & $(5.9\pm 3.2)\times 10^{-8}$     & $79$              & $<8.0\times 10^{-14}$         & \nod   & 12--13    \\
050925  & \nod           & $0.068\pm 0.027$ & $(7.5\pm 0.9)\times 10^{-8}$     & $100$             & $<3.0\times 10^{-14}$         & \nod   & 14--15    \\
051210  & $>1.55$        & $1.200\pm 0.200$ & $(8.1\pm 1.4)\times 10^{-8}$     & $840$             & $(2.4\pm 0.6)\times 10^{-12}$ & $-2.6$ & 16--17    \\
051221a & $0.5465$       & $1.400\pm 0.200$ & $(1.2\pm 0.1)\times 10^{-6}\,^c$ & $3.9\times 10^4$  & $(1.2\pm 0.1)\times 10^{-12}$ & $-1.2$ & 18--19    \\
051227  & \nod           & $8.000\pm 0.200$ & $(2.3\pm 0.3)\times 10^{-7}$     & $3.5\times 10^4$  & $(1.8\pm 0.6)\times 10^{-13}$ & $-1.1$ & 20--22    \\
060121  & \nod           & $1.970\pm 0.060$ & $(4.8\pm 0.3)\times 10^{-6}\,^d$ & $2.2\times 10^4$  & $(2.1\pm 0.5)\times 10^{-12}$ & $-1.2$ & 23--25    \\
060313  & $<1.7$         & $0.700\pm 0.100$ & $(1.1\pm 0.1)\times 10^{-6}\,^e$ & $1.0\times 10^5$  & $(1.5\pm 0.3)\times 10^{-13}$ & $-1.5$ & 26        \\
060502b & \nod$^f$       & $0.090\pm 0.020$ & $(4.0\pm 0.5)\times 10^{-8}$     & $2.8\times 10^4$  & $(2.4\pm 1.2)\times 10^{-14}$ & $-1.1$ & 10,17,27--29 \\
060801  & $1.1304$       & $0.500\pm 0.100$ & $(8.1\pm 1.0)\times 10^{-8}$     & $7.0\times 10^4$  & $<3.6\times 10^{-14}$         & \nod   & 10,17,30--31 \\
061006  & $0.4377$       & $0.420         $ & $(1.4\pm 0.1)\times 10^{-6}\,^g$ & $1.0\times 10^5$  & $(1.0\pm 0.4)\times 10^{-13}$ & $-1.9$ & 17,32     \\
061201  & $\sim 0.1\,^h$ & $0.800\pm 0.100$ & $(3.3\pm 0.3)\times 10^{-7}\,^i$ & $3.4\times 10^4$  & $(2.0\pm 0.5)\times 10^{-13}$ & $-1.9$ & 33--35    \\
061210  & $0.4095$       & $0.190         $ & $(1.1\pm 0.2)\times 10^{-6}\,^j$ & $2.3\times 10^5$  & $(1.4\pm 0.5)\times 10^{-13}$ & $-2.0$ & 17,36     \\
061217  & $0.8270$       & $0.212\pm 0.041$ & $(4.6\pm 0.8)\times 10^{-8}$     & $1340$            & $(4.7\pm 1.0)\times 10^{-13}$ & $-0.6$ & 17,37     \\
\enddata
\tablecomments{Prompt emission and X-ray Properties of the short GRB 
discussed in this paper, including (i) GRB name, (ii) redshift, (iii) 
duration, (iv) $\gamma$-ray fluence, (v) time of X-ray observations,
(vi) X-ray flux, (vii) X-ray temporal decay rate, and (viii) 
references.\\  
$^a$ Fluence is in the $15-150$ keV energy band unless otherwise
noted.\\  
$^b$ The fluence is in the $30-400$ keV band and for the initial
short pulse only; the total fluence including the $\sim 130$ s 
duration soft tail is $F_\gamma\approx 1.5\times 10^{-6}$ erg 
cm$^{-2}$ in the $2-400$ keV band \citep{vlr+05}.\\
$^c$ The fluence in the $20-2000$ keV band is $3.2\times 10^{-6}$
erg cm$^{-2}$ \citep{gcn4394}.\\
$^d$ The fluence is in the $2-400$ keV band.\\
$^e$ The fluence in the $20-2000$ keV band is $1.4\times 10^{-5}$
erg cm$^{-2}$ \citep{gcn4881}.\\
$^f$ A redshift of $z=0.257$ has been proposed by \citet{bpc+07} based
on a galaxy located $17.5''$ away from the XRT position, but a fainter
galaxy of unknown redshift has been identified within the error circle
\citep{bfp+06}.\\ 
$^g$ The fluence in the $20-2000$ keV band is $3.6\times 10^{-6}$
erg cm$^{-2}$ \citep{gcn5710}.\\
$^h$ The nature of the host is not clear, given the location of an
Abell cluster at $z=0.0865$ about $8'$ away \citep{gcn5880,gcn5995}, 
as well as a galaxy at $z=0.111$ about $17''$ away \citep{gcn5952}.\\
$^i$ The fluence in the $20-3000$ keV band is $5.3\times 10^{-6}$
erg cm$^{-2}$ \citep{gcn5890}.\\
$^j$ The fluence in the $100-1000$ keV band is $2.0\times 10^{-6}$
erg cm$^{-2}$ \citep{gcn5917}.\\
References: [1] \citet{gso+05}; [2] \citet{bpc+07}; [3]
\citet{vlr+05}; [4] \citet{ffp+05}; [5] \citet{hwf+05}; [6]
\citet{bcb+05}; [7] \citet{bpc+05}; [8] \citet{gbp+06}; [9]
\citet{gcn3793}; [10] \citet{nak07}; [11] \citet{ber06}; [12]
\citet{gcn3934}; [13] \citet{gcn3935}; [14] \citet{gcn4037}; [15]
\citet{gcn4043}; [16] \citet{lmf+06}; [17] \citet{bfp+06}; [18]
\citet{bgc+06}; [19] \citet{sbk+06}; [20] \citet{gcn4400}; [21]
\citet{gcn4401}; [22] \citet{gcn4402}; [23] \citet{dls+06}; [24]
\citet{ltf+06}; [25] \citet{pcg+06}; [26] \citet{rvp+06}; [27]
\citet{gcn5064}; [28] \citet{gcn5093}; [29] \citet{bpc+07}; [30]
\citet{gcn5381}; [31] \citet{gcn5382}; [32] \citet{sbp+06}; [33]
\citet{mps+06}; [34] \citet{gcn5952}; [35] \citet{gcn5995}; [36]
\citet{cbb+06}; [37] \citet{zbb+06}.}
\end{deluxetable}

\clearpage

\clearpage

\end{document}